\newcommand{\be}{\begin{equation}}
\newcommand{\ee}{\end{equation}}
\newcommand{\bea}{\begin{eqnarray}}
\newcommand{\eea}{\end{eqnarray}}
\newcommand{\ba}{\begin{array}}
\newcommand{\ea}{\end{array}}

\newcommand{\al}{\alpha}
\newcommand{\ga}{\gamma}
\newcommand{\Ga}{\Gamma}
\newcommand{\bet}{\beta}

\newcommand{\ka}{\kappa}
\newcommand{\de}{\delta}

\newcommand{\ep}{\epsilon}
\newcommand{\si}{\sigma}
\newcommand{\la}{\lambda}

\newcommand{\om}{\omega}
\newcommand{\Om}{\Omega}

\newcommand{\tha}{\theta}

\newcommand{\horava}{Ho\v{r}ava}
\newcommand{\EE}{E_{8}\times E_{8}}
\newcommand{\I}{S^{1}/{\bf Z}_{2}}
\newcommand{\tr}{{\rm tr}}
\newcommand{\pa}{\partial}
\newcommand{\rar}{\rightarrow}

\newcommand{\we}{\wedge}
\newcommand{\cF}{{\cal F}}
\newcommand{\cK}{{\cal K}}
\newcommand{\cZ}{{\cal Z}}

\documentclass[12pt,twoside]{article}

\begin{document}
 
\begin{flushright}
G\"{o}teborg ITP 98-01 \\ {\tt hep-th/9802148}
\end{flushright}
\vspace{1mm}
\begin{center}{\bf\Large\sf The heterotic prepotential from eleven dimensions}
\end{center}
\vskip 5mm
\begin{center}
Niclas Wyllard\footnote{wyllard@fy.chalmers.se} \vspace{5mm}\\
{\em  Institute of Theoretical Physics, S-412 96 G\"{o}teborg,
Sweden}
\end{center}
 
\vskip 5mm
 
\begin{abstract}
We compactify M-theory in the \horava{}-Witten formulation on $S^1/{\bf Z}_{2}\times K3\times T^2$. Focusing on the moduli-space of vector multiplets of the resulting four-dimensional $N=2$ theory, we determine the prepotential as an expansion in two dimensionless parameters which both scale as $\ka^{2/3}$. 
We determine the prepotential completely to relative order $\ka^{4/3}$ and compare the expression with the results obtained for the perturbative string theories. We find complete agreement to relative order $\ka^{4/3}$ between the strong and weak coupling regimes. The sources of higher order perturbative and non-perturbative corrections to the prepotential are also briefly discussed from the M-theory perspective.
\end{abstract}

\setcounter{equation}{0}
\section{Introduction and summary}
 
\horava{} and Witten conjectured that the strong coupling limit of the $\EE$ heterotic string is given by M-theory compactified on the line segment $I\cong S^1/{\bf Z}_{2}$ \cite{Horava:1996a}, and gathered convincing evidence which supported this claim \cite{Horava:1996a,Horava:1996b}. 
In particular, in \cite{Horava:1996b} they constructed the first few terms of the eleven-dimensional effective action. Witten has investigated other (partially) supersymmetry preserving configurations of M-theory, using the effective theory of ~\cite{Horava:1996b}. In \cite{Witten:1996a}, compactifications on the product of $I$ with a Calabi-Yau three-fold or with $K3$ were studied. 
Because of the presence of boundaries the dimensional reduction of the eleven-dimensional effective action becomes more intricate: in general it is not possible to take the fields to be independent of the eleventh coordinate and simultaneously satisfy the boundary conditions of the bulk fields. 
Following Witten's work several papers have appeared (starting with the works \cite{Banks:1996}) which consider the implications of the strongly coupled heterotic string for phenomenological aspects of $N=1$ vacua in four dimensions, see e.g. \cite{Nilles:1998} and references therein. 
In this paper we compactify M-theory in the \horava-Witten formulation on $S^1/{\bf Z}_{2}\times K3 \times T^2$ obtaining an effective four-dimensional theory with $N=2$ supersymmetry. We will focus on the moduli-space of vector multiplets of this theory. As is well known this moduli-space is subject to the restrictions of special K\"ahler geometry (for details, see e.g. \cite{Craps:1997}). The K\"ahler potential of a special K\"ahler manifold can be obtained from a holomorphic prepotential $\cF$ as 
\be
{\cal K}(\cZ^A,\bar{\cZ}^A) = - \log(i[2(\cF - \bar{\cF}) - (\cZ^B - \bar{\cZ}^B)(\cF_B + \bar{\cF}_B)])\, , 
\ee
where $\cF_{A} = \frac{\pa \cF}{\pa \cZ^A}$, and the bars denote complex conjugation. ${\cal Z}^A$ are the moduli --- the complex scalars of the vector multiplets. In our case the moduli are the geometric moduli arising from the internal manifold which together with the antisymmetric tensor moduli are organized into three complex scalars, conventionally denoted $S$, $T$ and $U$. 
We also have the moduli of the vector bundle over the torus. These will be denoted $V^{k}$, where $k=1\ldots 16$, and are often referred to as Wilson lines.

In the \horava-Witten formulation the $\EE$ vector multiplets arise as order $\ka^{2/3}$ corrections relative to the bulk supergravity action. This fact makes it natural to try to expand the effective action as a power series in $\ka^{2/3}$. 
In compactifications to four dimensions which preserve  $N=1$ supersymmetry, Choi et al \cite{Choi:1997} identified two natural dimensionless parameters, both of which were proportional to $\ka^{2/3}$, which could serve as expansion parameters of the four-dimensional K\"ahler potential. 
Adapted to our situation they are $\ep_{1}= \frac{\pi\rho \ka^{2/3}}{V_{K3}}$ and $\ep_{2}= \frac{\ka^{2/3}}{\pi\rho V_{T^2}}$, where $\pi\rho$ is the length of the interval, and $V_{K3}$, $V_{T^2}$ are the volumes of $K3$ and the torus, respectively. These parameters are the analogues of the world-sheet ($\ep_{\si}$) and string loop ($\ep_{s}$) expansion parameters appearing in the weakly coupled heterotic string. More precisely, as will be seen later, the expansion parameters can be written in terms of the moduli as
\bea
\ep_{1} &=& (4\pi)^{-1/3}\frac{{\rm Im}T}{{\rm Im}{S}}\,, \cr
\ep_{2} &=& (4\pi)^{-1/3}\frac{1}{{\rm Im}{T}} \,.
\eea 
In this paper we will calculate the K\"ahler potential (and from it obtain the associated prepotential) perturbatively as an expansion in these two dimensionless parameters. Since both parameters scale as $\ka^{2/3}$, where $\ka$ is the eleven-dimensional gravitational coupling constant, we will occasionally refer to the expansion simply as an expansion in $\ka^{2/3}$. In principle it is possible to form other dimensionless quantities which can be used as expansion parameters. 
However, in our case two parameters arise naturally since there are two moduli related to the sizes of the internal dimensions, namely $S$ and $T$, whereas the other moduli $U$ and $V^k$ are  related to the complex structure of the torus and the vector bundle over it and should not change when going from weak to strong coupling. 

We will assume that the torus and the K3 are isotropic and $V_{K3}^{1/4}\approx V_{T^2}^{1/2}$, and also that all relevant length scales are much larger that the eleven-dimensional Planck length $\ell_{11} \propto \ka^{2/9}$, i.e. $\pi\rho \gg \ell_{11}$, $V_{K3}^{1/4} \gg \ell_{11}$ and $V_{T^2}^{1/2}\gg \ell_{11}$. In addition we will also make a further assumption, namely $\pi\rho \gg V_{K3}^{1/4}\approx V_{T^2}^{1/2}$. 
This means that there is no region where the theory is effectively ten-dimensional. In terms of the moduli the above conditions translate into 
\be
({\rm Im} T)^2 \gg {\rm Im} S \gg (4\pi)^{-1/3}{\rm Im}T,\;\;\; {\rm Im}T\gg (4\pi)^{-1/3}\,,
\ee
where the last condition actually follows from the first two, and the first implies that $\ep_2 \ll \ep_1$.   
Witten showed \cite{Witten:1996a} that in order for the eleven dimensional description to make sense, $\ep_1$ was required to be small. 
We will see later that powers of $\ep_{1}$ are always accompanied by at least one power of $\ep_{2}$, hence, as noted in \cite{Choi:1997}, the expansion will work well even if $\ep_{1}$ is not small. As can be seen from the above expressions, ${\rm Im}S$ is bounded from below by $(4\pi)^{-2/3} \approx \frac{1}{5}$, hence we can not reach arbitrarily strong coupling. 
Choi et al suggested the following translation rules between the M-theory and perturbative heterotic string expansion parameters: $\ep_{1}^n\ep_{2}^m \sim \ep_{s}^n\ep_{\si}^{m + 2n}$, and conjectured that there will be a complete matching of the (perturbative) K\"ahler potentials in the two limits. 

The fact that the moduli are special coordinates, which means that they can not be arbitrarily changed, makes it plausible that it should be possible to extrapolate the prepotential of the weakly coupled theory into the strongly coupled regime, and that the entire ``perturbative'' prepotentials in the two limits should in fact agree. The relation of the prepotential to the spectrum of BPS states \cite{Harvey:1996a} gives further support to this idea.

We will determine the relative order $\ka^{2/3}$ and $\ka^{4/3}$ corrections to the prepotential completely in the situation where the size of the eleventh dimension is larger than the other compact dimensions, and all relevant length scales are larger than the Planck scale. We would like to stress that the calculation is a purely classical calculation.
Since the details of the calculation are rather technical we will state the result here. We find that for arbitrary number of Wilson lines and arbitrary instanton embeddings (with the proviso that there should be enough hyper multiplets present to make sufficient Higgsing possible) the prepotential to order $\ka^{4/3}$ (order $\ep_{1}\ep_{2}$) is
\be
^{2}\cF = S(TU - V^i V^i) \pm \frac{(k-12)}{2}TV^iV^i\,,
\ee
where the $\pm$ refers to the fact that we get different signs depending upon in which of the two $E_{8}$'s the Wilson lines are embedded, and $k$ is the instanton-number in the first $E_8$. 
Comparing this result with known results for the perturbative string theories with one Wilson line \cite{LopesCardoso:1997} we find a complete matching to order $\ka^{4/3}$ between the prepotentials in the two limits. This is further evidence for the correctness of the \horava-Witten conjecture.  
In the next section we will present the calculation of the terms in the prepotential of relative order $\ka^{0}$, $\ka^{2/3}$ and $\ka^{4/3}$. The final section is devoted to a discussion of how the other perturbative and non-perturbative corrections to the prepotential appear in the M-theory framework.

\setcounter{equation}{0}
\section{The perturbative prepotential}

In this section we will discuss the systematics of the $\ka^{2/3}$ expansion and determine the perturbative corrections to the prepotential up to and including the $\ka^{4/3}$ corrections.

\section*{$\ka^{0}$}

The bosonic fields of eleven-dimensional supergravity are the metric $g_{MN}$ and the antisymmetric 3-tensor $C_{IJK}$, with field strength $G_{IJKL} = \pa_{I}C_{JKL} \pm 23$ terms. 
The bosonic part of the eleven-dimensional supergravity action is to relative order $\ka^0$ 
\be
S_{11} = \frac{1}{2\ka^2}\int_{{\cal M}_{11}} \hspace{-2mm}d^{11}x\sqrt{-g}\left[ -R - \frac{1}{24}G_{IJKL}
G^{IJKL} - \frac{\sqrt{2}}{1728}\ep^{I_{1}\cdots I_{11}}C_{I_{1}I_{2}I_{3}}
G_{I_{4}I_{5}I_{6}I_{7}}G_{I_{8}I_{9}I_{10}I_{11}} \right] \,.\label{11daction}
\ee
Our conventions are as follows: capital roman indices take the values $0,\ldots,3,5,\ldots 11$, and are raised and lowered with the full eleven-dimensional metric. Lower case roman indices from the beginning of the alphabet label coordinates on $K3$, whereas greek indices refer to four-dimensions and are raised and lowered with the metric $g_{\mu\nu}$, defined in (\ref{11dmetric}) below. 
We will work on the manifold ${\cal M}_{11}={\cal M}_{4}\times T^2 \times K3\times S^1/{\bf Z}_{2}$ with boundaries at $x^{11}=0$ and $x^{11}=\pi\rho_0$. 
Because the manifold has boundaries we have to introduce corrections to the Einstein-Hilbert action in order to compensate for the variations of the Ricci-tensor localized at the boundaries.\footnote{From the Palatini identity $\sqrt{-g}g^{MN}\de R_{MN} = \pa_{I}(\sqrt{-g}g^{KI}\de\Ga^{L}_{KL} - \sqrt{-g}g^{KL}\de\Ga^{I}_{KL} )$, we see that the variation of the Ricci tensor gives rise to an unwanted boundary contribution, which has to be cancelled.} 
The corrections take the form \cite[appendix E.1]{Wald:1984}
\be
S^{K}_{10} = \frac{1}{\ka^2}[\int_{{\cal M}_{10}^{(1)}} d^{10}x \sqrt{-g_{10}}K^{(1)} +  \int_{{\cal M}_{10}^{(2)}} d^{10}\sqrt{-g_{10}}K^{(2)}]\,. \label{Kterms}
\ee
$K$ is the trace of the extrinsic curvature of the boundary, defined as $K_{IJ} = h_{I}^{\phantom{K}J}D_{K}n_{J}$, where $h_{MN}$ is the induced ten-dimensional metric: $h_{IJ} = g_{IJ} - n_{I}n_{J}$. The vector $n^{I}$ is a unit vector normal to the boundary. In our case $n_{I} = \pm \de_{I}^{11}\sqrt{g_{11,11}}$, which leads to $K  = g^{IJ}K_{IJ} = - h^{IJ}\Ga^{11}_{IJ}n_{11} = \frac{1}{2}g^{11,11}h^{IJ}n_{11}\pa_{11}g_{IJ}$.

To zeroth order (relative order $\ka^0$) we use a metric on ${\cal M}_{4}\times T^2 \times K3\times S^1/{\bf Z}_{2}$ of the form
\bea
ds_{11}^2 &=& e^{-4a-c-2b}g_{\mu\nu}dx^{\mu}dx^{\nu} + \frac{e^{2b}}{u_2}[dx^{5}dx^{5} + 2u_1dx^{5}dx^{6} + (u_1^2 + u_2^2)dx^{6}dx^{6}] \cr &+& e^{2a}\Om_{ab}dx^{a}dx^{b} 
+ e^{2c}dx^{11}dx^{11}\,, \label{11dmetric}
\eea
where $\Om_{ab}$ is a Ricci-flat K\"ahler metric on $K3$.
Inserting this expression into (\ref{11daction}) gives 
\bea
^{0}S_4 &=& \frac{\pi \rho_{0}V_{0}}{2\ka^2}\int d^{4}x\sqrt{-g}\Bigg[ -R - 
12\pa_{\mu}a\pa^{\mu}a 
- 8\pa_{\mu}a\pa^{\mu}b - 4\pa_{\mu}b\pa^{\mu}b - 4\pa_{\mu}a\pa^{\mu}c \cr 
&-& 2\pa_{\mu}b\pa^{\mu}c - \frac{3}{2}\pa_{\mu}c\pa^{\mu}c - \frac{1}{2}\frac{\pa_{\mu}u_1\pa^{\mu}u_1}{u_2^2}- \frac{1}{2}\frac{\pa_{\mu}u_2\pa^{\mu}u_2}{u_2^2} -\frac{1}{6}e^{8a + 4b}H_{\mu\nu\la}H^{\mu\nu\la}\cr &-& e^{-4b-2c}H_{\mu}H^{\mu} \Bigg]\,. \label{S41}
\eea
Here $H_{\mu\nu\la} = G_{\mu\nu\la 11}$ and $H_{\mu} = G_{\mu 5 6 11}$ are the zero modes of $G_{IJKL}$ relevant to our discussion. There are in general other components of $g_{MN}$ and $G_{IJKL}$ which give rise to scalars in four dimensions. However, these components lie partly in the $K3$ directions and give rise to scalars which will be shown below to belong to hyper multiplets, which are not considered in this paper.

In (\ref{S41}) $V_{0} =V^0_{K3}V^0_{T^{2}}$, where $V^0_{K3}$ and $V^{0}_{T^2}$ are the volumes in the metric without moduli factors, i.e. $V^{0}_{T^2}=\int_{T^2}d^2x$ and $V^{0}_{K3} = \int_{K3}d^4x \sqrt{\Om}$, where $\Om = \det{\Om_{ab}}$. Furthermore, $\pi\rho_0 = \int_{I}dx^{11}$.  The related physical (moduli dependent) volumes are: $V_{T^2}=e^{2b}V^{0}_{T^2}$, $V_{K3}=e^{4a}V^{0}_{K3}$, and $\pi\rho = e^{c}\pi\rho_0$.
  
The $K$ terms in the action (\ref{Kterms}) give no contribution at the current level, since $\pa_{11}g_{MN} =0$.
Introducing the new variables $\hat{a} := 4a + 2b$, $\hat{b} := 2b + c$, $\hat{c}
 := c + 2a$ and $U:= u_1 + i u_2$, and dualizing the 
three-form, by adding the term $-\frac{V_{0}\pi\rho_0}{6\sqrt{2}\ka^2}\int d^4x \sqrt{g}\si\ep^{\ga\mu\nu\la}\pa_{\ga}H_{\mu\nu\la}$ to the action, results in
\bea
^{0}S_4 &=& \frac{\pi \rho_{0}V_{0}}{2\ka^2}\int d^{11}x\sqrt{-g}\Bigg[ -R - 
\frac{1}{2}
\pa_{\mu}\hat{a}\pa^{\mu}\hat{a}  - \frac{1}{2}\pa_{\mu}\hat{b}\pa^{\mu}
\hat{b} - \pa_{\mu}\hat{c}\pa^{\mu}\hat{c}- \frac{1}{2}e^{-2\hat{a}}\pa_{\mu} \si\pa^{\mu} \si \cr &-&e^{-2\hat{b}}H_{\mu}H^{\mu} + 2\frac{\pa_{\mu}U
\pa^{\mu}\bar{U}}{(U-\bar{U})^{2}}\Bigg]\,.
\eea
If we define $S:= \si + ie^{\hat{a}}$ and $T:= \chi + ie^{\hat{b}}$, where 
$H_{\mu} = \frac{1}{\sqrt{2}}\pa_{\mu} \chi$, then ${^{0}}S_4$ can be written 
\bea
\hspace{-6mm} ^{0}S_4 \hspace{-2mm}&=& \hspace{-2mm}\frac{\pi \rho_{0}V_{0}}{2\ka^2}\int d^{11}x\sqrt{-g}\Bigg[ -R  - \pa_{\mu}\hat{c}
\pa^{\mu}\hat{c} + 2\frac{\pa_{\mu}S\pa^{\mu}\bar{S}}{(S-\bar{S})^{2}} + 
2\frac{\pa_{\mu}T\pa^{\mu}\bar{T}}{(T-\bar{T})^{2}}  + 2\frac{\pa_{\mu}U
\pa^{\mu}\bar{U}}{(U-\bar{U})^{2}}\Bigg]\,. \label{0S4}
\eea
The scalar $\hat{c}$ has to sit in a hyper-multiplet, since the kinetic terms  of the other scalars are derivable from a K\"ahler potential which in turn can be obtained from a prepotential, which implies that these scalars sit in vector multiplets. 
A more direct way of verifying that $\hat{c}$ sits in a hyper multiplet, is by first compactifying on $\I\times K3$ to six dimensions obtaining an $N=(1,0)$ supergravity theory coupled to Yang-Mills, with one massless antisymmetric tensor field $B_{ij}$, which means that we are discussing a vacuum with one tensor multiplet. 
In six dimensional $N=(1,0)$ theories the scalars sit in tensor multiplets (one scalar per multiplet) and hyper multiplets (4 scalars per multiplet). In the reduction we get, in addition to the scalars coming from the moduli of $K3$, two scalars: $\phi_{1} = 4a - c$ and $\phi_{2} =2\hat{c} = 4a + 2c$.  The  kinetic terms of these scalars arise from the reduction of the $R$ term in the action (\ref{11daction}) and are in the Einstein frame
\be
 S_{6} \propto \int d^6x\sqrt{-g_{6}}[- \pa_{i}\phi_{1} \pa^{i}\phi_{1} - \pa_{i}\phi_{2} \pa^{i}\phi_{2}] \,. 
\ee  
The scalar $\phi_{1}$ couples to the Yang-Mills field according to $S_{6}^{YM} \propto \int d^6x \sqrt{-g_{6}}e^{\frac{1}{2}\phi_{1}}\tr F_{ij}F^{ij}$. From the general structure of the couplings in $N=(1,0)$ supergravity \cite{Nishino:1986}, it follows that $\phi_{1}$ belongs to the tensor multiplet; thus, $\phi_{2}$ and all the other scalars have to be part of hyper multiplets. 
This continues to be true in four dimensions, since upon reduction on $T^2$ the hyper multiplets turn into four-dimensional hyper multiplets, and the tensor and vector multiplets give four-dimensional vector multiplets. 

In conclusion, to zeroth order in $\ka^{2/3}$ the action of the vector multiplets is derivable from the following prepotential and its associated K\"ahler potential
\bea
^{0}\cF = STU \;\; \Rightarrow&& ^{0}\cK = -\ka_{4}\ln(S-\bar{S})(T-\bar{T})(U-\bar{U})\,,
\eea
where $\ka_{4}=\frac{\pi\rho_{0}V_{0}}{\ka^{2}}$ is the four-dimensional gravitational coupling constant.

\section*{$\ka^{2/3}$}

The bosonic terms of order $\ka^{2/3}$ in the eleven-dimensional effective action of \horava-Witten  are localized at the ten-dimensional boundaries and are given by\footnote{Capital roman indices with bars take the values $0,\ldots,3,5,\ldots 10$ and are raised with the induced ten-dimensional metric.}
\bea
S^{YM}_{10} &=& -\frac{1}{16\pi \ka^2}\left( \frac{\ka}{4\pi}\right)^{2/3}\int_{{\cal M}_{10}^{(1)}} d^{10}x \sqrt{-g_{10}} [\tr F^{(1)}_{\bar{M}\bar{N}}F^{(1)\bar{M}\bar{N}} - \frac{1}{2}e_4(R^{(1)})]  \cr &&-\frac{1}{16\pi \ka^2}\left( \frac{\ka}{4\pi}\right)^{2/3}\int_{{\cal M}_{10}^{(2)}} d^{10}x \sqrt{-g_{10}} [\tr F^{(2)}_{\bar{M}\bar{N}}F^{(2)\bar{M}\bar{N}}- \frac{1}{2}e_4(R^{(2)})]\,,
\eea
where
\be
e_4(R) = R_{\bar{I}\bar{J}\bar{K}\bar{L}}R^{\bar{K}\bar{L}\bar{I}\bar{J}} - 4R_{\bar{I}\bar{J}}R^{\bar{I}\bar{J}} + R^{2}\,, \label{R2}
\ee  
and the superscripts ${}^{(1)}$ and ${}^{(2)}$ refer to the two boundaries situated at $x^{11} = 0$ and $x^{11} = \pi\rho_{0}$, respectively. 
  Evidence for the necessity of the $R^2$ terms in the action above was given by Lukas et al \cite{Lukas:1997b}. We will find their presence crucial for consistency (e.g. with the results obtained in \cite{Witten:1996a}). 
In the above action we have also taken into account the numerical alterations discussed in \cite{Conrad:1997}. However, we have not be able to exclude the possibility that the original numerical factors of \cite{Horava:1996b} are correct.

There are two sources of order $\ka^{2/3}$ corrections to the four-dimensional action. First, we get corrections from the boundary action above by inserting the zeroth order metric. The ten-dimensional metric is obtained by restricting the eleven-dimensional metric (\ref{11dmetric}) to the ten-dimensional hyper-planes, i.e.
\bea
ds^{2}_{10} &=& e^{-4a-c-2b}g_{\mu\nu}dx^{\mu}dx^{\nu} + \frac{e^{2b}}{u_2}(dx^{5}dx^{5} + 2u_1dx^{5}dx^{6} + (u_1^2 + u_2^2)dx^{6}dx^{6}) \cr &+& e^{2a}\Om_{ab}dx^{a}dx^{b} \,,
\eea
where to this order the induced metrics are the same on both boundaries.
Inserting this expression into the above  boundary action leads to
\be
 -\frac{V_{0}}{16\pi \ka^2}\left( \frac{\ka}{4\pi}\right)^{2/3}\int d^{4}x \sqrt{-g} \frac{e^{-\hat{b}}}{u_2}\Big[ 2(u_1^2 + u_2^2)\pa_{\mu}A^{k}_{5}\pa^{\mu}A^{k}_{5} + 2\pa_{\mu}A^{k}_{6}\pa^{\mu}A^{k}_{6} - 4u_1\pa_{\mu}A^{k}_{5}\pa^{\mu}A^{k}_{6}\Big]\,, \label{A20}
\ee
where $k$ is summed over appropriate values, and $A^k_{5,6}$ are the $U(1)$ gauge potential on the $T^2$. One has to be careful about the $R^2$ and $F^2$ terms associated with $K3$, because of the topological restrictions and the requirements of anomaly cancelation. In general these terms gives a contribution to the four-dimensional effective action. 
However, at the current level this contribution vanishes as a consequence of the anomaly cancelation requirement that the total instanton number in the two $E_{8}$'s should equal 24, the Euler characteristic of $K3$, i.e. 
\be
 \frac{1}{16\pi^2}\int_{K3} (F^{(1)}\we F^{(1)} + F^{(2)}\we F^{(2)}) = \frac{1}{16\pi^2}\int_{K3} R\we R\,,
\ee
which translates into
\be
 \frac{1}{32\pi^2}\int_{K3} \sqrt{\Om}(F^{(1)}_{ab}F^{(1)ab} + F^{(2)}_{ab}F^{(2)ab}) = \frac{1}{32\pi^2}\int_{K3}\sqrt{\Om} e_{4}(R)\,,
\ee
since the instanton field strengths and the curvature of $K3$ are self-dual.
 Recalling that $T= \chi + ie^{\hat{b}}$ we see that (\ref{A20}) can be rewritten as
\be
 -\frac{V_{0}}{8\pi \ka^2}\left( \frac{\ka}{4\pi}\right)^{2/3}\int d^{4}x \sqrt{-g} \frac{2i}{T-\bar{T}}\frac{2i}{U-\bar{U}}\Big[ (\pa_{\mu} A^{k}_{6} - U\pa_{\mu}A^{k}_{5})(\pa^{\mu} A^{k}_{6} - \bar{U}\pa_{\mu}A^{k}_{5})\Big]\,.
\ee
The second source of $\ka^{2/3}$ terms arise through deformations of the zeroth order eleven-dimensional action (\ref{11daction}). Because of the presence of sources, localized on the ten-dimensional boundaries, in the equations of motions for the bulk fields and in the Bianchi identity for $G_{IJKL}$, it is in general not possible to take $G_{IJKL}$ and $g_{MN}$ to be independent of $x^{11}$, as in conventional dimensional reduction. 
We will expand $g_{MN}$ and $G_{IJKL}$ as power series in $\ka^{2/3}$. We write \be
g_{MN} = g_{MN}^{({\rm zm})} + {^{1}}g_{MN}+ {\cal O}(\ka^{4/3}),\;\; G_{IJKL} = G^{({\rm zm})}_{IJKL} + {^{1}}G_{IJKL} + {\cal O}(\ka^{4/3})\,,
\ee
 where in general the corrections depend on $x^{11}$, and use a self-consistent method to derive expressions for the deformations. For our purposes it is sufficient to determine $G_{IJKL}$ to first order in four-dimensional derivatives and $g_{MN}$ to second order. 
The method we use involves solving the equations of motion perturbatively and substituting the result back into the action order by order \cite{Lukas:1997b}. This procedure is equivalent to integrating out non-trivial massive modes as discussed in \cite{Lukas:1997c}. We will start by determining the deformations which are zeroth order in four-dimensional derivatives. 
Witten \cite{Witten:1996a} constructed a supersymmetric configuration, preserving one-half of the supersymmetries, of eleven-dimensional supergravity on ${\cal M}_{6}\times K3\times S^1/{\bf Z}_2$, to all orders in a perturbative expansion in $\ka^{2/3}$. 
This solution is not expected to remain valid in the full M-theory, since it will in general be modified by higher order terms in the M-theory effective action, however it is sufficient for our purposes. In the approximation where the size of the eleventh dimension is much larger than the size of $K3$ Witten's solution for $G_{IJKL}$ is, in our normalizations:
\bea
G^{(W)}_{abcd} &=& \frac{2\pi\sqrt{2}}{V^{0}_{K3}}\left(\frac{\ka}{4\pi}\right)^{2/3}(k-12)\ep_{abcd} \cr
G^{(W)}_{abc11} &=& 0\,. \label{G0W}
\eea 
Here $k$ is the instanton number in the first $E_8$. The conditions for unbroken supersymmetry then determine \cite{Witten:1996a} the metric to be\footnote{We have taken into account the results of \cite{Dudas:1997}.}
\be
ds_{11}^2 = (1 + \sqrt{2}w)^{-1/3}g_{ij}dx^{i}dx^{j} + (1 + \sqrt{2}w)^{2/3}(\Om_{ab}dx^adx^b + dx^{11}dx^{11})\,. \label{gW}
\ee
Here $\ep_{abcd}$ is the $\ep$-tensor in the $\Om_{ab}$ metric and $w = -\frac{2\pi\sqrt{2}}{V^{0}_{K3}}\left(\frac{\ka}{4\pi}\right)^{2/3}(k-12)(x^{11}-\pi\rho_0)$. 
As we will see later the modification, due to the introduction of moduli fields,  of Witten's original solution is
\bea
G^{(W)}_{abcd} &=& \frac{2\pi\sqrt{2}}{V^{0}_{K3}e^{4a}}\left(\frac{\ka}{4\pi}\right)^{2/3}(k-12)\ep_{abcd} \cr
G^{(W)}_{abc11} &=& 0 \,,\label{GW}
\eea
and the first order correction to the metric becomes
\be
g^{(W)}_{MN} = - e^{c-4a}\frac{8\pi}{3V^{0}_{K3}}\left( \frac{\ka}{4\pi}\right)^{2/3}(x^{11} - \frac{\pi\rho_0}{2})(k-12)g^{(0)}_{MN}\,, \label{gW1a}
\ee
when $M,N =7,\ldots,11$ and
\be
g^{(W)}_{MN} = + e^{c-4a}\frac{4\pi}{3V^{0}_{K3}}\left( \frac{\ka}{4\pi}\right)^{2/3}(x^{11} - \frac{\pi\rho_0}{2})(k-12)g^{(0)}_{MN}\,, \label{gW1b}
\ee
when $M,N =0,\ldots,3,5,6$. 
In the above expressions, $g^{(W)}_{MN}$ has been chosen to satisfy $\langle g^{(W)}_{MN} \rangle_{x^{11}} = 0$, where $\langle\cdots\rangle_{x^{11}} := \int_{0}^{\pi\rho_0}\cdots$.
In general, the splitting of the metric into different parts is ambiguous. However, by redefining the zero-mode part $g^{({\rm zm})}_{IJ}$ by an $x^{11}$ independent (but in general $\ka$ dependent) term in such a way that the average of each of the other parts over the orbifold interval vanishes, the ambiguity can be removed. 

Next we turn to the
relative order $\ka^{2/3}$ deformations of $G_{IJKL}$ which are first order in four-dimensional derivatives.
The equation of motion obtained by varying the action, together with the Bianchi identity for $G_{IJKL}$ are\footnote{Our conventions are as in \cite{Horava:1996b}, i.e. $F_{[\bar{I}\bar{J}}F_{\bar{K}\bar{L}]} = \frac{1}{4!}(F_{\bar{I}\bar{J}}F_{\bar{K}\bar{L}} \pm 23$ terms$)$, $G_{IJKL} = \pa_{I}C_{JKL} \pm 23$ terms, and $(dG)_{IJKLM} = \pa_{I}G_{JKLM} +$ cyclic permutations.}
\bea
\frac{1}{\sqrt{g}}\pa^{I}(\sqrt{g}G_{IJKL}) &=& \frac{\sqrt{2}}{1152}\ep_{JKLI_{1}\cdots I_{8}}G^{I_{1}\cdots I_{4}}G^{I_{5}\cdots I_{8}} \cr
(dG)_{11\bar{I}\bar{J}\bar{K}\bar{L}} &=& -\frac{3}{\sqrt{2}\pi}\left( \frac{\ka}{4\pi}\right)^{2/3}\Big[ \de(x_{11})(\tr F^{(1)}_{[\bar{I}\bar{J}}F^{(1)}_{\bar{K}\bar{L}]} - \frac{1}{2}\tr R^{(1)}_{[\bar{I}\bar{J}} R^{(1)}_{\bar{K}\bar{L}]}) \cr &+& \de(x_{11} - \pi\rho)(\tr F^{(2)}_{[\bar{I}\bar{J}}F^{(2)}_{\bar{K}\bar{L}]} - \frac{1}{2}\tr R^{(2)}_{[\bar{I}\bar{J}} R^{(2)}_{\bar{K}\bar{L}]})\Big] \,. \label{eqm}
\eea

Ho\v{r}ava and Witten showed that the Bianchi identity for $G_{IJKL}$ with the $\de$-function source terms is equivalent to $dG=0$ together with the boundary conditions 
\bea
G_{\bar{I}\bar{J}\bar{K}\bar{L}}|_{x^{11}=0} &=& -\frac{3}{2\sqrt{2}\pi}\left( \frac{\ka}{4\pi}\right)^{2/3}(\tr F^{(1)}_{[\bar{I}\bar{J}}F^{(1)}_{\bar{K}\bar{L}]} - \frac{1}{2}\tr R^{(1)}_{[\bar{I}\bar{J}} R^{(1)}_{\bar{K}\bar{L}]}) \cr
G_{\bar{I}\bar{J}\bar{K}\bar{L}}|_{x^{11}=\pi\rho} &=& \frac{3}{2\sqrt{2}\pi}\left( \frac{\ka}{4\pi}\right)^{2/3}(\tr F^{(2)}_{[\bar{I}\bar{J}}F^{(2)}_{\bar{K}\bar{L}]} - \frac{1}{2}\tr R^{(2)}_{[\bar{I}\bar{J}} R^{(2)}_{\bar{K}\bar{L}]})\,.
\eea
These two alternative formulations were referred to in \cite{Horava:1996b} as the upstairs and downstairs approaches, respectively.  In the downstairs picture one works on the manifold with boundaries ${\cal M}_{10}\times \I$, with non-trivial boundary conditions on the bulk fields. In the upstairs method one instead works on the manifold ${\cal M}_{10}\times S^1$ with $\de$-function sources at the fixed points of the ${\bf Z}_{2}$ action. In this paper we will almost exclusively use the downstairs approach. 
In order to solve the above equations we start by neglecting the $G^2$ terms in the equation of motion. We thus want to solve $dG = 0$ and $D^{I}G_{IJKL} = 0$, supplemented by the above boundary conditions. 
In the case where the sources vary slowly over the size of the orbifold interval the solution to the above boundary-value problem is \cite{Lukas:1997b} 
\be
C^{(B)}_{\bar{I}\bar{J}\bar{K}} = -\frac{1}{4\sqrt{2}\pi}(\frac{\ka}{4\pi})^{2/3}\Big[\om^{(1)} - \left(\frac{x^{11}}{\pi\rho_0}\right)(\om^{(1)} + \om^{(2)})\Big]_{\bar{I}\bar{J}\bar{K}}\,,
\ee
where $\om = \om^{YM} - \frac{1}{2}\om^{L}$ and $\om^{YM}_{\bar{K}\bar{L}\bar{M}} = A_{\bar{K}}(\pa_{\bar{J}}A_{\bar{L}} - \pa_{\bar{L}}A_{\bar{J}}) + \frac{2}{3}A_{\bar{K}}[A_{\bar{J}},A_{\bar{L}}] +$ cyclic permutations, is the Yang-Mills Chern-Simons form. Furthermore, $\pa_{\bar{I}}\om^{YM}_{\bar{K}\bar{L}\bar{M}} \pm$ cyclic permutations $= 6F_{[\bar{I}\bar{J}}F_{\bar{K}\bar{L}]}$, which leads to 
\bea
G^{(B)}_{\bar{I}\bar{J}\bar{K}\bar{L}} &=& -\frac{3}{2\sqrt{2}\pi}(\frac{\ka}{4\pi})^{2/3}\Big[\tr F^{(1)}_{[\bar{I}\bar{J}}F^{(1)}_{\bar{K}\bar{L}]} - \frac{1}{2}\tr R^{(1)}_{[\bar{I}\bar{J}} R^{(1)}_{\bar{K}\bar{L}]} \cr &-& \left(\frac{x^{11}}{\pi\rho_0}\right)\bigg\{\tr F^{(1)}_{[\bar{I}\bar{J}}F^{(1)}_{\bar{K}\bar{L}]} - \frac{1}{2}\tr R^{(2)}_{[\bar{I}\bar{J}} R^{(2)}_{\bar{K}\bar{L}]}\bigg\} \Big] \cr
G^{(B)}_{\bar{I}\bar{J}\bar{K}11} &=& -\frac{1}{4\sqrt{2}\pi^2\rho_{0}}(\frac{\ka}{4\pi})^{2/3}\Big[\om^{(1)} + \om^{(2)}\Big]_{\bar{I}\bar{J}\bar{K}}\,. \label{GB}
\eea
As an aside we see that we recover Witten's solution (\ref{G0W}), with the moduli-dependence given in (\ref{GW}), by turning on the fields on the $K3$ and integrating $G^{(B)}_{abcd}$ given in (\ref{GB}) over $K3$ using  the following normalizations of the instantons 
\bea
\int_{K3} d^4x\sqrt{\Om}F_{ab}F^{ab} &=& 2 \int_{K3} F\we *F =  2 \int_{K3}F\we F =  2\cdot16\pi^2 \frac{p_{1}(V)}{2} = 32\pi^{2}m, \;\;m\in {\bf Z} \cr
\int_{K3} d^4x\sqrt{\Om}e_{4}(R) &=& 2 \int_{K3} R\we *R =  2 \int_{K3}R\we R =  2 \cdot16\pi^2 \frac{p_{1}(K3)}{2} = 32\pi^{2} 24\,, \label{norms}
\eea
where $F = \frac{1}{2}F_{ab}dx^{a}\we dx^{b}$. We also have $\int_{K3}d^4x\sqrt{\Om}\ep^{abcd}F_{ab}F_{cd} = 64\pi^2m$.
 We will be interested in the case where, apart from the non-trivial gauge-fields on $K3$ required for anomaly cancelation, only the $U(1)$ gauge potentials $A^k_5$ and $A^k_6$ are non-zero. This leads to, if we look at the case with $\{ \bar{I}\bar{J}\bar{K}\bar{L} \} = \{\mu 56 11\}$ 
\be
G^{(B)}_{\mu 56 11} = \frac{1}{4\sqrt{2}\pi^2\rho_{0}}\left(\frac{\ka}{4\pi}\right)^{2/3}\Big[A^{k}_{5}\pa_{\mu}A^{k}_{6} - A^{k}_{6}\pa_{\mu}A^{k}_{5}\Big]\,.
\ee
This is the only contribution to this order if we keep only the part of $G^{(B)}_{IJKL}$ that is linear in four-dimensional derivatives. However, this is not the end of the story, since we have to take the $G^2$ terms in the equation of motion into account. The above expressions for $G^{(W)}_{IJKL}$ induces corrections to the equation of motion. Inserting the results for the corrections which are zeroth order in four-dimensional derivatives into the equations of motion gives 
\bea
\frac{1}{\sqrt{g}}\pa^{11}(\sqrt{g}G^{(H)}_{11\mu\nu\la}) &=& 2\frac{\sqrt{2}}{1152}\ep^{(0)}_{\mu\nu\la\si 5611abcd}G^{({\rm zm})\si5611}G^{(W)abcd} \cr
\frac{1}{\sqrt{g}}\pa^{11}(\sqrt{g}G^{(H)}_{11\mu56}) &=& 2\frac{\sqrt{2}}{1152}\ep^{(0)}_{\mu56\nu\la\si 11abcd}G^{({\rm zm})\nu\la\si 11}G^{(W)abcd}\,,
\eea
which in turn leads to 
\bea
G^{(H)}_{\mu\nu\la 11} &=& \frac{2\pi}{V^{0}_{K3}}\left(\frac{\ka}{4\pi}\right)^{2/3}(x^{11} - \frac{\pi\rho_{0}}{2})(k-12)[\frac{4}{3}e^{c-4a}H_{\mu\nu\la} - 2e^{-8a-4b}\ep_{\mu\nu\la\si}H^{\si}]     \cr
G^{(H)}_{\mu 56 11} &=& \frac{2\pi}{V^{0}_{K3}}\left(\frac{\ka}{4\pi}\right)^{2/3}(x^{11} - \frac{\pi\rho_{0}}{2})(k-12)[\frac{4}{3}e^{c-4a}H_{\mu} - \frac{1}{3}e^{4b+2c}\ep_{\mu\nu\la\si}H^{\nu\la\si}] \,,
\eea
where $\ep_{\mu\nu\la\si}$ is the $\ep$-tensor in the $g_{\mu\nu}$ metric. Furthermore, we have also imposed the condition $\langle G_{\bar{I}\bar{J}\bar{K}11}^{(H)} \rangle_{x^{11}} =0$. From this property it follows that the $G^{2}$ terms involving $G^{(H)}_{IJKL}$ will not give any contribution to the four-dimensional effective action at order $\ka^{2/3}$.

The final result for the first order correction is $^{1}G_{IJKL} = G^{(W)}_{IJKL} + G^{(B)}_{IJKL} + G^{(H)}_{IJKL}$, which solves the equations of motion and satisfies the Bianchi identity, up to corrections which contains more than one four-dimensional derivative. 

The next step is to determine the first order correction to the metric. 
Einstein's equations resulting from the terms in the eleven dimensional action up to order $\ka^{2/3}$ are in the {\em upstairs} picture \cite{Lukas:1997b}
\be
R_{IJ} + \frac{1}{6}(G_{IKLM}G_{J}^{\phantom{J}KLM} - \frac{1}{12}g_{IJ}G^{2}) =-\frac{1}{2\pi}\left( \frac{\ka}{4\pi}\right)^{2/3}[\de(x_{11})S_{IJ}^{(1)} + \de(x_{11}- \pi\rho)S_{IJ}^{(2)}] \,,
\ee
where the sources are given by the expressions 
\bea
S_{\bar{I}\bar{J}}^{(i)} &=& \frac{1}{\sqrt{g_{11,11}}}\left[\tr (F^{(i)})_{\bar{I}\bar{K}}(F^{(i)})_{\bar{J}}^{\phantom{J}\bar{K}} - \frac{1}{12}g_{\bar{I}\bar{J}}\tr (F^{(i)})^{2}\right] \cr
S_{11,11}^{(i)} &=& \frac{1}{6}\sqrt{g_{11,11}}\,\tr (F^{(i)})^{2}\,.
\eea
We again observe that the $G^2$ terms cannot be neglected to order $\ka^{2/3}$. We write $^{1}g_{IJ} = g^{(W)}_{IJ} + g^{(B)}_{IJ} + g^{(H)}_{IJ}$, where $g^{(W)}_{IJ}$ is the metric determined by Witten (\ref{gW1a}), (\ref{gW1b}), i.e. the deformation resulting from the sources coming from the $K3$ part. 
Similarly, $g^{(B)}_{IJ}$ is the contribution induced by the background Yang-Mills fields on the $T^2$, and $g^{(H)}_{IJ}$ is the part induced by the $G^2$ terms. Assuming that the size of the eleventh dimension is much greater than the other compact dimensions, and keeping only terms with no more than two four dimensional derivatives, the linearized Ricci tensor becomes 
\bea
R_{\bar{M}\bar{N}} &=& ^{0}R_{\bar{M}\bar{N}} - \frac{1}{2}{^{0}}g^{({\rm zm})11,11}\pa_{11}^{2}g_{\bar{M}\bar{N}}^{(B+H)}  \cr
R_{11,11} &=& ^{0}R_{11,11} - \frac{1}{2}{^{0}}g^{({\rm zm})\bar{K}\bar{L}}\pa_{11}^{2}g_{\bar{K}\bar{L}}^{(B+H)}\,,
\eea
We have the gauge freedom to impose the harmonic coordinate conditions $g^{MN}\Ga^{K}_{MN} = 0$, which to first order imply that  $^{1}g_{11,11} = {^{0}}g^{({\rm zm})}_{11,11} {^{1}}g_{\bar{K}\bar{L}}{^{0}}g^{({\rm zm})\bar{K}\bar{L}} + \ga$, where $\ga$ is independent of $x^{11}$, and can be set to zero by redefining $g^{({\rm zm})}_{11,11}$. From these results it follows that $R^{(B+H)}_{MN} = - \frac{1}{2}g^{({\rm zm})11,11}\pa_{11}^{2}g_{MN}^{(B+H)}$. 

We will follow the method of \cite{Lukas:1997b,Lalak:1997a} and solve the linearized Einstein equations as an expansion in (four-dimensional) derivatives. 
We start by  neglecting the $G^{2}$ terms and assume that $g^{(B)}_{MN}$ is second order in four-dimensional derivatives. 
We also neglect derivatives with respect to the $K3$ and $T^2$ directions on the left hand side of Einstein's equations, but not the $K3$ dependence on the right hand side since the sources are dimension four operators and thus can be ``absorbed'' by integration over $K3$. This is not possible for higher order operators which thus can be neglected.  
A more convincing way to treat the dependence on the $K3$ coordinates is perhaps to use Witten's solution and not bother about the fields on $K3$ (however, it is important to check that the two methods give the same results); either way one can drop all derivatives except those in the $x^{11}$ direction on the left hand side. Neglecting the $G^2$ terms  we can reformulate the problem in the downstairs picture as \cite{Lukas:1997b} 
\bea
\pa_{11}\pa_{11}g^{(B)}_{MN} &=& \frac{e^{2c}}{2\pi^2\rho_0}\left(\frac{\ka}{4\pi}\right)^{2/3}[S^{(1)}_{MN} + S^{(2)}_{MN}] \cr
\pa_{11}g^{(B)}_{MN}|_{x^{11}=0} &=& -e^{2c}\frac{1}{2\pi}\left(\frac{\ka}{4\pi}\right)^{2/3}S^{(1)}_{MN} \cr 
\pa_{11}g^{(B)}_{MN}|_{x^{11}=\pi\rho_0} &=& e^{2c}\frac{1}{2\pi}\left(\frac{\ka}{4\pi}\right)^{2/3}S^{(2)}_{MN}\,,
\eea
where we have used the particular form of the zeroth order affine connection to replace $D_{11}$ with $\pa_{11}$.
The general solution to the above boundary-value problem subject to the requirement $\langle g^{(B)}_{MN}\rangle_{x^{11}}=0$ is
\be
g^{(B)}_{MN} = e^{2c}\frac{1}{2\pi}\left(\frac{\ka}{4\pi}\right)^{2/3}\left[ \left(\frac{(x^{11})^2}{2\pi\rho_0} - x^{11} + \frac{\pi\rho_0}{3}\right)S^{(1)}_{MN} + \left(\frac{(x^{11})^2}{2\pi\rho_0} - \frac{\pi\rho_0}{6}\right)S^{(2)}_{MN}\right]\,. \label{gB}
\ee
In the above discussion we have not included the $R^{2}$ terms; these are needed in order to correctly reproduce Witten's result (\ref{gW1a}), (\ref{gW1b}) for the metric, by integrating $g_{IJ}$ given in (\ref{gB}) over $K3$, with the field strengths on the $K3$ turned on, and using (\ref{norms}).

The first order deformation of the metric arising from the Wilson lines is given in (\ref{gB}), where the only non-zero components of $F_{MN}$ are $F_{\mu 5}$ and $F_{\mu 6}$. 
Finally, the $G^2$ terms in Einstein's equation gives rise to $H_{\mu}$ and $H_{\mu\nu\la}$ dependent corrections to $g_{MN}$, e.g.
\bea
g^{(H)}_{\mu\nu} &=& \frac{4\pi e^{2c}}{3V^0_{K3}}\left(\frac{\ka}{4\pi}\right)^{2/3}\left[ \frac{1}{6}(x^{11}-\frac{\pi\rho_0}{2})^3 - \frac{(\pi\rho_0)^2}{2}(x^{11}-\frac{\pi\rho_0}{2})\right](k-12) \cr
&& \hspace{-5mm}\times \Bigg\{[\frac{4}{3}e^{4a+4b+c}H_{\mu\la\si}H_{\nu}^{\phantom{\nu}\la\si} - 2H_{\mu\la\si}\ep_{\nu}^{\phantom{\nu}\la\si\bet}H_{\bet} - \frac{1}{3}H_{\mu}\ep_{\nu}^{\phantom{\nu}\la\si\bet}H_{\la\si\bet} +\frac{4}{3}e^{-4a-4b-c}H_{\mu}H_{\nu}] \cr &&-\frac{1}{12}g_{\mu\nu}[\frac{4}{3}e^{4a+4b+c}H_{\ga\la\si}H^{\ga\la\si} - \frac{7}{3}H_{\ga\la\si}\ep^{\ga\la\si\bet}H_{\bet} +\frac{4}{3}e^{-4a-4b-c}H_{\ga}H^{\ga}]\Bigg\}\,.
\eea
We will now discuss the corrections to the four-dimensional action arising from the first order deformations of $g_{MN}$ and $G_{IJKL}$. 
Using the above results we see that we get an order $\ka^{2/3}$ correction from the $G^{(0)}_{\mu5611}G^{(B)\mu5611}$ term in the action (\ref{11daction}), viz. 
\be
 -\frac{1}{\ka^2}\frac{V_{0}}{8\pi}\left( \frac{\ka}{4\pi}\right)^{2/3}\int d^{4}x \sqrt{-g} e^{-2\hat{b}}\Big[\pa_{\mu}\chi(A^{k}_{5}\pa^{\mu}A^{k}_{6} - A^{k}_{6}\pa_{\mu}A^{k}_{5})\Big]\,.
\ee
Furthermore, we get corrections from the $R$ term in the action by inserting ${^{1}}g_{MN}$ --- the first order correction to the metric. These take the form 
\bea
\hspace{-10mm}&&-\frac{2}{3}\Bigg\{ -\frac{1}{\ka^2}\frac{V_{0}}{8\pi}\left( \frac{\ka}{4\pi}\right)^{2/3}\int d^{4}x \sqrt{-g} e^{-2\hat{b}}\Big[\pa_{\mu}\chi(A^{k}_{5}\pa^{\mu}A^{k}_{6} - A^{k}_{6}\pa^{\mu}A_{k}^{5})\Big] \cr \hspace{-10mm}&&- \frac{V_{0}}{8\pi \ka^2}\left( \frac{\ka}{4\pi}\right)^{2/3}\int d^{4}x \sqrt{-g} \frac{2i}{T-\bar{T}}\frac{2i}{U-\bar{U}}\Big[ (\pa_{\mu} A^{k}_{6} - U\pa_{\mu}A^{k}_{5})(\pa^{\mu} A^{k}_{6} - \bar{U}\pa^{\mu}A^{k}_{5})\Big] \Bigg\}\,.
\eea
Note that although $\langle g^{(B)}_{MN}\rangle_{x^{11}} = 0$, it does not necessarily follow that $\langle \sqrt{-g}R\rangle_{x^{11}} = 0$, since $R$ contains derivatives with respect to $x^{11}$. We also get contributions from the $K$ terms on the boundary, which by construction exactly cancel the contributions coming from $\sqrt{-g}R$. Combining the above results we get 
\bea
^{1}S_{4} &=& \frac{V_{0}}{2 \pi \ka^2}\left( \frac{\ka}{4\pi}\right)^{2/3}\int d^{4}x \sqrt{-g} \frac{1}{T-\bar{T}}\frac{1}{U-\bar{U}}\Big[ (\pa_{\mu} A^{k}_{6} - U\pa_{\mu}A^{k}_{5})(\pa^{\mu} A^{k}_{6} - \bar{U}\pa^{\mu}A^{k}_{5})\Big]  \cr &+&\frac{V_{0}}{2 \pi \ka^2}\left( \frac{\ka}{4\pi}\right)^{2/3}\int d^{4}x \sqrt{-g} \frac{1}{(T-\bar{T})^{2}}\Big[\pa_{\mu}\chi (A^{k}_{5}\pa^{\mu}A^{k}_{6} - A^{k}_{6}\pa^{\mu}A^{k}_{5})\Big]\, .
\eea
At order $\ka^{0}$ the normalization of $S$ and $T$ was left undetermined. We fix the normalization of $S$ by demanding that the four-dimensional gauge sector arising from turning on $F_{\mu\nu}$, should be described by
\be
S^{S}_{\rm gauge} = -\frac{1}{16\pi}\int d^{4}x \sqrt{-g}\, {\rm Im}(S)[ \tr(F^{(1)}_{\mu\nu}F^{(1)\mu\nu}) + \tr(F^{(2)}_{\mu\nu}F^{(2)\mu\nu})]\,, \label{g}
\ee
which implies that $S = \frac{\tha}{2\pi} + i\frac{4\pi}{g^{2}}$. From (\ref{g}) we get the associated instanton action $- 2\pi {\rm Im}S = -\frac{8\pi^2}{g^2}$. The convention (\ref{g}) leads to  $S = V_{0}\left( \frac{1}{4\pi\ka^{2}}\right)^{2/3}[ \si + ie^{\hat{a}}]$. We will see later that this is a natural normalization of $S$. Similarly, corrections at order $\ka^{4/3}$ determine the normalization of $T$. If we demand
\be 
S^{T}_{\rm gauge} = -\frac{(k-12)}{2}\frac{1}{16\pi}\int d^{4}x \sqrt{-g}\, {\rm Im}(T)[ \tr(F^{(1)}_{\mu\nu}F^{(1)\mu\nu}) - \tr(F^{(2)}_{\mu\nu}F^{(2)\mu\nu})] \,,
\ee
we obtain $T = \rho_{0} V^{0}_{T^2}\left( \frac{1}{4\pi\ka^2}\right)^{1/3}[ \chi + ie^{\hat{b}}]$. Notice that the gauge coupling constants of the two $E_{8}$'s are positive as a consequence of the earlier given bounds on ${\rm Im}S$ and ${\rm Im}T$. We also define \cite{LopesCardoso:1994} 
\be
V^{k} := \frac{(V^{0}_{T^2})^{1/2}}{4\pi}(A_{6}^{k} - U A_{5}^{k})\,.
\ee 
We have to redefine $T$ in order for it to remain a special coordinate as 
\be
T = \rho_{0} V^{0}_{T^2}\left( \frac{1}{4\pi\ka}\right)^{2/3}[ \chi + ie^{\hat{b}}] + \frac{V^{0}_{T^2}}{(4\pi)^2}A_{5}^{k}(A_{6}^{k} - U A_{5}^{k})\,.
\ee
discussedUsing these definitions the above action can at this order be obtained from the following prepotential and its associated K\"ahler potential
\bea
\hspace{-8mm}^{1}\cF = S[TU - V^{k}V^{k}] \Rightarrow && \hspace{-5mm} ^{1}\cK = -\ka_{4}\ln[(S-\bar{S})((T-\bar{T})(U-\bar{U}) - (V^k-\bar{V}^k)^{2})]\,.
\eea
At relative order $\ka^{2/3}$ it is actually not necessary to assume that the size of the eleventh dimension is much larger than the other compact directions, this means that the result is independent of the ratio between the two sizes and thus should agree with the result obtained in the perturbative heterotic string. However, as we will see later at higher orders in $\ka^{2/3}$, even when the eleventh dimension is larger than the other internal dimensions the prepotentials will agree. The large amount of supersymmetry present is the main reason for this correspondence, as discussed in the introduction. 

\section*{$\ka^{4/3}$}

We are now in a position to start our investigations of the $\ka^{4/3}$ corrections. First of all we will assume that only integer powers of $\ka^{2/3}$  appear in the ``perturbative'' expansion of the effective action. 
One way to motivate this assumption is the following \cite{Russo:1997}: upon compactification of M-theory from eleven to ten dimensions the ten-dimensional expansion parameter $\al '$ is related to $\ka$ as $\al' \propto \ka^{2/3}$. In addition, similar to the fact that $\al '$ is related to the inverse string tension, in eleven dimensions $\ka^{2/3}$ is related to the membrane tension. 
In general, other powers of the eleven dimensional Planck-length $\ka^{2/9}$ could appear in the expansion since the natural cut-off is proportional to $\ka^{2/9}$. Even if other powers appear they are not expected to interfere with our calculation to order $\ka^{4/3}$, but might become important at higher orders. On the boundary there are no dimensionally allowed terms involving $R$, $F$ and $G$ which could enter at order $\ka^{4/3}$. In the bulk on the other hand, terms like $G^{8}$, $G^{6}R$, $G^{4}R^{2}$, $G^{2}R^{3}$ and $R^{4}$ are allowed by dimensional counting. However, these terms are either higher order than two in four-dimensional derivatives or higher order in $\ka^{2/3}$, hence they can be neglected at the order we are presently working. 

We will start by calculating the terms in the effective four dimensional action induced by the $\ka^{2/3}$ corrections to $G_{IJKL}$ and $g_{MN}$. The contribution from the Chern-Simons term is\footnote{Since an explicit $C_{IJK}$ appears in this term we really need at least some of the corrections to $G_{IJKL}$ which are second order in four-dimensional derivatives. However, by using the Bianchi identity together with integrations by parts, we get an expression involving only the parts of $G_{IJKL}$ ($C_{IJK}$) which are first order in four dimensional derivatives.} 
\be
 \frac{\sqrt{2}}{24\ka^2}\pi\rho_0 V^{0}_{T^2} \left(\frac{\ka}{4\pi}\right)^{4/3}(k-12)\int d^4 x \sqrt{-g}\ep^{\mu\nu\la\si}H_{\mu\nu\la}(\om^{(1)}_{YM} - \om^{(2)}_{YM})|_{\si56}
\ee
where $\ep^{\mu\nu\la\si}$ is the $\ep$-tensor in the $g_{\mu\nu}$ metric.
As the next step we will calculate the deformation of the Yang-Mills action on the boundary. Since the eleven dimensional metric at order $\ka^{2/3}$ depends on $x^{11}$ the induced ten-dimensional metrics will in general be different on the two boundaries. The result is 
\bea
&&-\frac{\pi\rho_{0}V^{0}_{T^2}}{4\ka^2}(k-12)\left( \frac{\ka}{4\pi}\right)^{4/3}\hspace{-1mm}\int d^{4}x \sqrt{-g} \frac{e^{-\hat{a}}}{u_2}\Big[ (\pa_{\mu} A^{(1)k}_{6} - U\pa_{\mu}A^{(1)k}_{5})(\pa^{\mu} A^{(1)k}_{6} - \bar{U}\pa_{\mu}A^{(1)k}_{5}) \cr && \hspace{10mm}- (\pa_{\mu} A^{(2)k}_{6} - U\pa_{\mu}A^{(2)k}_{5})(\pa^{\mu} A^{(2)k}_{6} - \bar{U}\pa_{\mu}A^{(2)k}_{5})\Big]\ \label{FT2}
\eea
We also get non-zero corrections from the parts involving the non-trivial gauge-fields and Riemann curvature on the $K3$,  since at this order the contributions from the two boundaries add. Using the normalizations (\ref{norms}) and inserting $g^{(B)}_{MN}$ we get a contribution which exactly equals (\ref{FT2}). Furthermore,  by inserting $g^{(W)}_{\bar{M}\bar{N}}$ using $g^{(W)}_{\bar{M}\bar{N}}|_{x^{11}=0} = -g^{(W)}_{\bar{M}\bar{N}}|_{x^{11}=\pi\rho_0}$, we get
\be
8\pi^2\frac{\pi\rho_0V^{0}_{T^2}}{\ka^2V^{0}_{K3}}(k-12)^2\left( \frac{\ka}{4\pi}\right)^{4/3}\int d^4x\sqrt{-g}e^{-12a-c-2b} \label{pot}
\ee
This term has to be cancelled by terms of the same ilk coming from the $R$, $K$ and $G^2$ terms, since a potential is forbidden by the requirements of $N=2$ supersymmetry. This is a non-trivial check of our conventions and normalizations. The contribution to the potential coming from the $R$, $G^2$ and $K$ terms is
\be
(\frac{4}{3} - 4 -\frac{16}{3})\pi^2\frac{\pi\rho_0V^{0}_{T^2}}{\ka^2V^{0}_{K3}}(k-12)^2\left( \frac{\ka}{4\pi}\right)^{4/3}\int d^4x\sqrt{-g}e^{-12a-c-2b}\,,
\ee
which indeed cancels (\ref{pot}). In fact, the potential vanishes to all orders in Witten's supergravity solution as required. Moreover, when inserted into the $R$ term of the action $g^{(W)}_{MN}$ does not give any corrections to the zeroth order action (\ref{0S4}), which are second order in four-dimensional derivatives. This is also true to all orders in Witten's supergravity solution. 

The contributions arising from inserting $g^{(B)}_{\bar{M}\bar{N}}$ (and $g^{(W)}_{IJ}$) into the $R$ and  $K$ terms, become
\bea
&&(-\frac{19}{36} + \frac{7}{9} )\frac{\pi\rho_{0}V^{0}_{T^2}}{\ka^2}(k-12)\left( \frac{\ka}{4\pi}\right)^{4/3}\int d^{4}x \sqrt{-g} \frac{e^{-\hat{a}}}{u_2} \Bigg[ (\pa_{\mu} A^{(1)k}_{6} - U\pa_{\mu}A^{(1)k}_{5}) \cr &&\times(\pa^{\mu} A^{(1)k}_{6} - \bar{U}\pa^{\mu}A^{(1)k}_{5}) - (\pa_{\mu} A^{(2)k}_{6} - U\pa_{\mu}A^{(2)k}_{5})(\pa^{\mu} A^{(2)k}_{6} - \bar{U}\pa^{\mu}A^{(2)k}_{5})\Bigg]
\eea
Finally, the only other contribution comes from $G^{(B)}_{\mu5611}G^{(B)\mu5611}$ and is
\be
-\frac{1}{\ka^2}\frac{V_{0}}{64\pi^3\rho_0}\left( \frac{\ka}{4\pi}\right)^{4/3}\int d^{4}x\sqrt{-g}e^{-2\hat{b}}[A_{5}^{k}\pa_{\mu}A_{6}^{k} - A_{6}^{k}\pa_{\mu}A_{5}^{k}]^{2} \,.
\ee
This term is already captured by the previous prepotential. 
Actually, as we will see later, the above results comprise the total contribution to the four-dimensional effective action.  

There are other contributions which we have glossed over so far. These are of the form
\bea
&&\frac{\pi^2}{2 \ka^2}\frac{(\pi\rho_0)^3V^{0}_{T^2}}{V^{0}_{K3}}(k-12)^2\left( \frac{\ka}{4\pi}\right)^{4/3}\int d^4x\sqrt{-g}\Bigg[ \al e^{4b+2c}H_{\mu\nu\la}H^{\mu\nu\la} + \beta e^{-8a-4b}H_{\mu}H^{\mu} \cr &&+ \ga e^{c-4a}\ep^{(0)}_{\mu\nu\la\si}H^{\mu\nu\la}H^{\si} \Bigg]
\eea
and arise from inserting, $G^{(H)}_{IJKL}$, $g^{(H)}_{MN}$ and $g^{(W)}_{MN}$ into the various terms of the eleven-dimensional action. However, from the fact that we did not get any corrections to the action involving (four-dimensional) derivatives of the real parts of the $S$ and $T$ moduli, we do not expect any corrections involving the imaginary parts of $S$ and $T$. Thus the terms of the above form 
should sum to zero (possibly after some field redefinitions).

We now turn to the ${\cal O}(\ka^{4/3})$ corrections to $G_{IJKL}$. It will turn out that these corrections will not lead to any contributions to the four-dimensional action at ${\cal O}(\ka^{4/3})$, but we will include a brief discussion for completeness. First we note that the expressions (\ref{GW}) and (\ref{GB}) still satisfy the Bianchi identity. (\ref{GW}) also satisfies the equation of motion and hence is valid unchanged to ${\cal O}(\ka^{4/3})$. The equations of motion for $G_{\mu\nu\la 11}$ and $G_{\mu5611}$ gets corrections as before. The solution becomes 
\bea
^{2}G_{\mu\nu\la 11} &=& d_1\frac{e^c}{V^{0}_{K3}e^{4a}\pi\rho_0}\left(\frac{\ka}{4\pi}\right)^{4/3}(k-12)[x^{11}- \frac{\pi\rho_0}{2}]\ep^{(0)}_{\mu\nu\la\si}(A^k_5\pa^{\si} A^k_6 - A^k_6\pa^{\si} A^k_5) \cr
&+& (\frac{(k-12)}{V^{0}_{K3}})^{2}\left(\frac{\ka}{4\pi}\right)^{4/3}[\frac{(x^{11})^2}{2} - \frac{\pi\rho_{0}}{2}x^{11} + \frac{(\pi\rho_0)^2}{12}] \cr &&\times e^{c-4a}[d_2e^{c-4a}H_{\mu\nu\la} + d_3 e^{-8a-4b}\ep^{(0)}_{\mu\nu\la\si}H^{\si}]  \cr
^{2}G_{\mu 56 11} &=& \tilde{d}_1\frac{e^c}{V^{0}_{K3}e^{4a}\pi\rho_0}\left(\frac{\ka}{4\pi}\right)^{4/3}(k-12)[x^{11} -\frac{\pi\rho_0}{2}](A^k_5\pa_{\mu} A^k_6 - A^k_6\pa_{\mu} A^k_5)  \cr
&+& (\frac{(k-12)}{V^{0}_{K3}})^{2}\left(\frac{\ka}{4\pi}\right)^{4/3}[\frac{(x^{11})^2}{2} - \frac{\pi\rho_{0}}{2}x^{11} + \frac{(\pi\rho_0)^2}{12}] \cr &&\times e^{c-4a}[\tilde{d}_{2}e^{c-4a}H_{\mu} + \tilde{d}_{3}e^{4b+2c}\ep^{(0)}_{\mu\nu\la\si}H^{\nu\la\si}] \,,
\eea
where the $d_{i}$'s and $\tilde{d}_{i}$'s are constants. From the above formula we see that these terms will not contribute to the four-dimensional action at order $\ka^{4/3}$.
In principle it is possible to determine the part of $G_{IJKL}$ which is first order in derivatives to ``all'' orders, i.e. the solution of the equation of motion given in  (\ref{eqm}) using the ``all'' order results for $g^{(W)}_{MN}$ and $G^{(W)}_{IJKL}$; it is possible to write down a differential equation from which $G_{IJKL}$ can be obtained. However, the equations of motion are expected to get corrections from higher order terms in the full M-theory effective action.

We will not need the second order corrections to the metric, since the contribution from the $R$ term will be cancelled by the contribution coming from $K$. However, in principle it is possible to calculate these corrections.
 
In conclusion, the $\ka^{4/3}$ corrections to the action are 
\bea
^{2}S_4 &=& 2\frac{(k-12)}{2}\frac{\pi \rho_{0}V_{0}}{\ka^2}\frac{V_{T^2}}{(4\pi)^2}\int d^{11}x\sqrt{-g}\Bigg[ \frac{1}{(S-\bar{S})^{2}}\Big[\pa_{\mu}\si (A^{(1)k}_{5}\pa^{\mu}A^{(1)k}_{6} - A^{(1)k}_{6}\pa_{\mu}A^{(1)k}_{5})\Big] \cr &&+\frac{1}{S-\bar{S}}\frac{1}{U-\bar{U}}\Big[ (\pa_{\mu} A^{(1)k}_{6} - U\pa_{\mu}A^{(1)k}_{5})(\pa^{\mu} A^{(1)k}_{6} - \bar{U}\pa_{\mu}A^{(1)k}_{5})\Big] - {^{(1)}} \leftrightarrow {^{(2)}} \Bigg] \,.
\eea 

After redefining $S$ according to $S \rar V_{0}\left( \frac{1}{4\pi\ka^{2}}\right)^{2/3}[ \si + ie^{\hat{a}}] \mp \frac{V^{0}_{T^2}}{(4\pi)^2}A_{5}^{k}(A_{6}^{k} - U A_{5}^{k})$, the K\"ahler potential can be obtained from the following prepotential (valid to order $\ka^{4/3}$) 
\be
^{2}\cF = S(TU - V^2) \pm \frac{(k-12)}{2}TV^2\,, \label{F2}
\ee
where $\pm$ refers to the fact that depending upon in which $E_{8}$ the Wilson lines are embedded we get different signs. The above expression should be compared with the prepotential in the weakly coupled heterotic string. In the case with one Wilson line embedded in the first $E_{8}$ the prepotential has been discussed in \cite{LopesCardoso:1997} based on earlier results obtained using mirror symmetry \cite{Berglund:1997}. 
If we compare (\ref{F2}) with the result in \cite{LopesCardoso:1997} with $n$ replaced by $k-12$, we see that the two expressions for the prepotential agree to order $\ka^{4/3}$. We would like to stress that our results apply to all instanton embeddings (at least at level $\ka^{4/3}$) and arbitrary number of Wilson lines (with the proviso that there should be enough hyper multiplets present to make sufficient Higgsing possible).

We will end this section by briefly discussing the M-theory corrections to the minimal eleven-dimensional supergravity action. In the bulk and on the boundary a $G^{2n}$ term with at most two four-dimensional derivatives is at least of order $(\ka^{2/3})^{2n-2}$ whereas a $R^n$ term is at least of order $(\ka^{2/3})^{2n-2}$ because the part of $R_{IJKL}$ with no four-dimensional derivatives is zero to order $\ka^{2/3}$ since $g^{(W)}_{IJ}$ is linear in $x^{11}$. By dimensional counting $G^4$, and $G^2 R$ terms are possible at order $\ka^{2/3}$ on the boundary. 
These terms, if present, will possibly affect the four-dimensional K\"ahler potential at order $\ka^{6/3}$. In ten dimensions it is known that the $R^2$ terms, in order to be consistent with supersymmetry, have to be evaluated with a shifted spin connection $\tilde{\om}^{AB}_{\bar{K}} = \om^{AB}_{\bar{K}} + \frac{1}{\sqrt{2}}H^{AB}_{\bar{K}}$ \cite{Bergshoeff:1989}. Here $H^{AB}_{\bar{K}} = e^{A}_{\bar{I}}e^{B}_{\bar{J}}H^{\bar{I}\bar{J}}_{\bar{K}}$, where $e^A_{\bar{I}}$ is the zehn-bein.
This will probably continue to be true in eleven dimensions and takes care of the $G_{\bar{I}\bar{J}\bar{K}11}=H_{\bar{I}\bar{J}\bar{K}}$ terms. The corrections arising from this modification all have more than two derivatives in the four-dimensional sense. 
  In the bulk the dimensionally allowed terms at order $\ka^{4/3}$, e.g. the $R^4$ terms, will possibly contribute to the four-dimensional K\"ahler potential at order $\ka^{16/3}$.

\setcounter{equation}{0}
\section{Higher order and non-perturbative effects}

In this section we will discuss higher order  and non-perturbative corrections to the prepotential calculated in the previous section.
The (heterotic) prepotential has the following general structure
\be
{\cal F} = S(TU - V^2) + f_{0}(T,U,V) + \sum_{n=1}^{\infty}f_{n}(T,U,V)e^{2\pi n i S}\,. \label{prep}
\ee
This structure follows from the discrete Peccei-Quinn symmetry $S\rar S+1$, which is believed to be unbroken even in the strongly coupled regime. The restriction on the summation in the last term follows from the fact that the sum should be well behaved in the weakly coupled limit $S\rar i\infty$. In the perturbative regime $|q|=|e^{2\pi iS}|$ is small, however the expansion should not be thought of as a weak coupling expansion; rather, its structure is fixed by the Peccei-Quinn symmetry. Furthermore, even at fairly strong coupling $|q|$ is smaller than 1, so there is no indication that the series should diverge. In a suitable Weyl chamber 
$f_{0}(T,U,V)$ in the equation above can be written
\be
f_{0}(T,U,V) = TQ(U,V) + g_{0}(U,V) + \sum_{n=1}^{\infty}g_{n}(U,V)e^{2\pi n i T}\,, \label{f0}
\ee
where $Q(U,V)$ is a quadratic form in $U$ and the $V^k$'s. The restriction on $f_{0}(T,U,V)$ follows from the symmetry (unbroken at this level) $T\rar T+1$ together with the fact that the prepotential should be well behaved on the entire moduli-space, in particular when $T\rar i \infty$ (the decompactification limit of $T^2\times S^{1}/{\bf Z}_{2}$). The form of the above perturbative prepotential implies that $f_{0} \rar f_{0} + Q$ under $T\rar T+1$. 

From the above expressions together with the fact that $S$ is of order $\ka^{-4/3}$ and $T$ is of order $\ka^{-2/3}$ we see that the other terms in the exact weakly coupled perturbative prepotential of \cite{LopesCardoso:1997} are either at least of order $\ka^{6/3}$ or even ``non-perturbative'', relative to the $STU$ part. 

How does the $g_0(U,V)$ term, which is known to be of the form $\sum c_{l,b}e^{(lU + b_{k} V^k)}$ arise in the eleven dimensional picture? Consider the eleven-dimensional theory compactified to eight dimensions on $\I\times T^2$. Besides corrections of the previous type, there is also another effect which has to be taken into account. (Finite) supergravity one-loop corrections  to the kinetic terms (obtained from integrating out KK modes) depending on the moduli of the torus, i.e. $T$, $U$ and $V$ are expected to occur at order $\ka^{6/3}$. Upon reduction to four dimensions these terms will on general grounds be of the form 
\be
\int d^4x \sqrt{-g} \frac{h(U,\bar{U},V,\bar{V})}{(T-\bar{T})(S-\bar{S})}[\frac{\pa_{\mu}U\pa^{\mu}\bar{U}}{(U-\bar{U})^2} + \frac{\pa_{\mu}T\pa^{\mu}\bar{T}}{(T-\bar{T})^2} + \ldots]\,.
\ee
Notice that this term is of relative order $\ka^{6/3}$. In order for it to be derivable from a prepotential, $h$ must satisfy a differential equation, which when $V^k=0$ is $\frac{\pa}{\pa U}\frac{\pa}{\pa\bar{U}}\frac{h(U,\bar{U})}{U-\bar{U}} = 2(U - \bar{U})^{2}\frac{h(U,\bar{U})}{U-\bar{U}}$, and in addition be invariant under the $SL(2,{\bf Z})$ symmetry of the torus. Taken together these requirements almost completely determine $h$, in cases with a low number of moduli.
We expect instanton corrections to $h(U,\bar{U},V,\bar{V})$ coming from euclidean membranes suspended between the two boundaries and wrapped on the $T^2$. The action for these instanton is -$T_2 V_{T^2}\pi\rho_0$, where $T_2$ is the tension of the M-theory membrane. 
If we use the expressions for the tensions of the membrane and five-brane of M-theory as given e.g. in \cite{Conrad:1997} i.e. $T_{2} = 2\pi\left(4\pi\ka^2\right)^{-1/3}$ and $T_{5}=2\pi(4\pi\ka^2)^{-2/3}$, together with the normalization of ${\rm Im}T$ given in the previous section, we see that the instanton action is $-2\pi{\rm Im}T$ which is the correct normalization, cf. (\ref{f0}). 
These instanton modifications are similar to the modification of the $R^4$ terms in type II theories, see e.g \cite{Green:1997d} and references therein.  The structure of the corrections is constrained by invariance under U-duality.
It might be interesting to investigate the structure of $f_0(T,U,V)$ from this vantage point in more detail.
 
How do the rest of the terms in the prepotential (\ref{prep}) arise from the M-theory perspective? These terms are instanton corrections which are invisible in the perturbative heterotic string. In the M-theory picture these corrections are obtained by wrapping euclidean five-branes on $K3 \times T^2$  similarly to what was done in \cite{Harvey:1996b}. 
As in \cite{Harvey:1996b} we can perform a consistency check of our normalization of $S$. The instanton action of a euclidean five-brane wrapped on $K3\times T^2$ is $-T_5V_{0}$, where $T_5$ is the five-brane tension. 
If we use the expressions for the tension given above together with the formula for the ten-dimensional gauge coupling constant $\la^2 = 4\pi(4\pi\ka^{2})^{2/3}$, we get an instanton action of the form $-T_5V_{0} = -\frac{8\pi^2 V_0}{\la^2}= -\frac{8\pi^2}{g^2}$, where $g$ is the four-dimensional gauge coupling constant, i.e. the same result as in the previous section. The two ways we have wrapped the branes of M-theory to get the instanton corrections are the only two configurations of completely wrapped branes which preserve supersymmetry \cite{Lalak:1997a}.

In this paper we have considered  theories with one tensor multiplet in the six-dimensional sense. More general cases with more than one ``$S$-like'' modulus in four dimensions fall outside the scope of this paper.

Since a lot of information about the exact expression for the (perturbative) prepotential is known in certain cases \cite{Harvey:1996a}, it is conceivable that it is possible to make use of the information in lower dimensions in reverse to obtain information about the structure of the higher order corrections to the effective action of M-theory. 
Another challenge is to try to obtain information about the non-perturbative $f_{n}$'s in (\ref{prep}), possibly using U-duality invariance together with the automorphic properties \cite{Berglund:1997b} of the $f_n$'s. 

\section*{Acknowledgements}
I am grateful to M\aa ns Henningson for stimulating discussions and comments. I would also like to thank Bengt Nilsson for discussions, and the theory division at CERN for hospitality during the completion of this work.

\def\href#1#2{#2}


\begingroup\raggedright\endgroup

\end{document}